\documentclass[12pt,oneside]{article}
\usepackage{amsfonts,amssymb,graphicx}
\usepackage{cite}
\usepackage{indentfirst}
\usepackage{newlfont}
\usepackage{amsthm}
\usepackage{amsmath}
\usepackage{latexsym}
\usepackage{eucal}
\usepackage{eufrak}
\usepackage{amssymb}
\usepackage{mathrsfs}
\usepackage{graphicx}

\newtheorem{proposition}{Proposition}

\newtheorem{lemma}{Lemma}

\newtheorem{corollary}{Corollary}

\DeclareMathOperator{\card}{card}

\begin{document}

\begin{center}
\vspace{1truecm}
{\bf\Large Rigorous Results on the Bipartite Mean-Field Model}\\
\vspace{1cm}
{Micaela Fedele $^{\dagger}$ and Francesco Unguendoli $^{\ddagger}$ }\\
\vspace{.5cm}
{\small $^{\dagger}$ Dipartimento di Matematica, Universit\`a di Bologna,
 e-mail: {\em micaela.fedele2@unibo.it}}\\ 
 \vspace{.5cm}
 {\small $^{\ddagger}$ Dipartimento di Matematica Pura e 
Applicata, Universit\`a di Modena e Reggio Emilia, e-mail: {\em francesco.unguendoli@unimore.it}}\\
\end{center}

\vskip 1truecm
\begin{abstract}\noindent
We consider a bipartite mean-field model in which both the interaction constant and the external field take different
values only depending on the groups particles belong to. We compute the exact value of the thermodynamic limit 
of the model exploiting a tail estimation on the number of configurations that share the same value of the magnetization and 
we analyze the critical points of the pressure functional associated to the symmetric version of the model as the external 
field is away or small.

\end{abstract}

\section*{Introduction}
Bipartite mean-field model have been
introduced since the 50s to reproduce the phase transition of some materials
called metamagnets. In particular in 
\cite{gorter1956transitions}, \cite{bidaux1967antiferromagnetisme} and \cite{kincaid1975phase}
a bipartite mean-field model is used to approximate a two-sublattice with
nearest neighbor and next-nearest neighbor exchanging interactions. The
same model has also been used to study the loss of gibbsianess for a system
that evolves according to a Glauber dynamics 
\cite{külske2007spin}.

In recent times the general version of these models have been proposed in
the attempt to describe the large scale behaviour of some socio-economic
systems 
\cite{contucci2007modeling}, assuming that individual’s decisions depends upon the decisions of others.

The investigation of the model introduced in \cite{contucci2007modeling} has been
pursued at a mathematical level in \cite{gallo2008bipartite}. It has been shown the existence
of the thermodynamic limit of the pressure exploiting a monotonicity condition on the Gibbs state of the Hamiltonian (see \cite{bianchi2003thermodynamic}). The factorization
of the correlation functions has been proved for almost every choice of parameters and the exact solution of the thermodynamic limit is computed
whenever the Hamiltonian is a convex function of the magnetizations.

In this paper we carry on the analysis of the mathematical property of the bipartite mean field model.
Firstly we
compute the exact solution of the thermodynamic limit. We exploit a tail
estimation on the number of configurations that share the same value of the
vector of the magnetization. This technique is the same used by Talagrand
to compute the thermodynamic limit for the Curie-Weiss model \cite{talagrand2003spin}. Then we analyze the critical points of the pressure functional
associated to a symmetric bipartite mean-field model in the case in which the external field is absent or small. This analysis highlights for which values of parameters the model undergoes a
phase transition.

This work is organized as follows. Chapter one introduces the notations and states
the main results. Chapter 2 contains the proofs. 
The appendix contains the proof of the
lemmas that make the paper self contained.

\section{Definition and Statement}
We consider a system of $N$ particles that can be divided into $2$ subsets $P_{1},P_{2}$ with $P_{1}\cap P_{2}=\emptyset$,
and sizes $|P_{l}|=N_{l}$, where $N_{1}+N_{2}=N$. Particles interact 
with each other and with an external field according to the mean field Hamiltonian:
\begin{equation}\label{hamiltoniana.1}
H_{N}(\boldsymbol{\sigma})=-\frac{1}{2N}\sum_{i,j=1}^{N}J_{ij}\sigma_{i}\sigma_{j}-\sum_{i=1}^{N}h_{i}\sigma_{i}.
\end{equation}
The $\sigma_{i}$ represents the spin of the particle $i$, while $J_{ij}$ is the parameter that tunes the mutual 
interaction between the particle $i$ and the particle $j$ and takes values according to the following symmetric matrix:

\begin{displaymath}
         \begin{array}{ll}
                \\
                N_1 \left\{ \begin{array}{ll}
                \\
                                  \end{array}  \right.
                \\
                N_2 \left\{ \begin{array}{ll}
                \\
            \\
                \\
                                  \end{array}  \right.
         \end{array}
          \!\!\!\!\!\!\!\!
         \begin{array}{ll}
                \quad
                 \overbrace{\qquad }^{\textrm{$N_1$}}
                 \overbrace{\qquad \qquad}^{\textrm{$N_2$}}
                  \\
                 \left(\begin{array}{c|ccc}
                               \mathbf{J}_{11}  &  & \mathbf{ J}_{12}
                                \\
                                 \hline
                                &  &  &
                                \\
                                \mathbf{ J}_{12}^{t} &  & \mathbf{ J}_{22}
                                \\
                                &   &   &
                                \\
                      \end{array}\right)
               \end{array}
\end{displaymath}\\
\noindent where each block $\mathbf{J}_{ls}$ has constant elements $J_{ls}$. For $l=s$, $\mathbf{J}_{ll}$ is a square matrix, whereas the matrix $\mathbf{ J}_{ls}$ is rectangular and all entries can be either positive or negative allowing both ferromagnetic and anti-ferromagnetic interactions. The vector field takes also different values depending on the subset the particles belong to as specified by: 

\begin{displaymath}
         \begin{array}{ll}
                N_1 \left\{ \begin{array}{ll}
                                      \\
                                   \end{array}  \right.
                                        \\
                N_2 \left\{ \begin{array}{ll}
                                        \\
                                          \\
                                         \\
                 \end{array}  \right.
           \!\!\!\!\!\!
    \end{array}
    \!\!\!\!\!\!
    \left(\begin{array}{ccc|c}
                \mathbf{h}_{1}
            \\
            \hline
            \\
            \mathbf{h}_{2}
            \\
            \\
        \end{array}\right)
\end{displaymath}\\
\noindent where each $\mathbf{h}_{l}$ is a vector of constant elements $h_{l}$.
The joint distribution of a spin configuration $\boldsymbol{\sigma}=(\sigma_{1},\dots ,\sigma_{N})$ is given by the 
Boltzmann-Gibbs measure:
\begin{equation}\label{misura.BG}
P_{N,\mathbf{J},\mathbf{h}}\{\boldsymbol{\sigma}\}=\frac{\exp(-\beta H_{N}(\boldsymbol{\sigma}))}{Z_{N}(\mathbf{J},\mathbf{h})}
\end{equation}
where $Z_{N}(\mathbf{J},\mathbf{h})$ is the partition function
\begin{equation*}
Z_{N}(\mathbf{J},\mathbf{h})=\sum_{\boldsymbol{\sigma}}\exp(-\beta H_{N}(\boldsymbol{\sigma})).
\end{equation*}
By introducing the magnetization of a set of spins $A$ as:
\begin{equation*}
m_{A}(\boldsymbol{\sigma})=\frac{1}{|A|}\sum_{i \in A}\sigma_{i}
\end{equation*}
\noindent and indicating by $m_{l}(\boldsymbol{\sigma})$ the magnetization of the set $P_{l}$, and by $\alpha_{l}=N_{l}/N$ 
the relative size of the set $P_{l}$, we may easily express the Hamiltonian (\ref{hamiltoniana.1}) as:
\begin{equation}\label{hamiltoniana.2}
 H_{N}(\boldsymbol{\sigma})=-Ng\Big(m_{1}(\boldsymbol{\sigma}),m_{2}(\boldsymbol{\sigma})\Big)
\end{equation}
\noindent where the function $g$:
\begin{align}\label{funzione.g}
g(m_{1}(\boldsymbol{\sigma}),m_{2}(\boldsymbol{\sigma}))&=\frac{1}{2}\bigg(\alpha_{1}^{2}J_{11}m_{1}(\boldsymbol{\sigma})^{2}+2\alpha_{1}\alpha_{2}J_{12}m_{1}(\boldsymbol{\sigma})m_{2}(\boldsymbol{\sigma})+
\alpha_{2}^{2}J_{22}m_{2}(\boldsymbol{\sigma})^{2}\bigg)+\nonumber\\
&\quad+\alpha_{1}h_{1}m_{1}(\boldsymbol{\sigma})+\alpha_{2}h_{2}m_{2}(\boldsymbol{\sigma})
\end{align}
 depends from the reduced interaction matrix $\mathbf{J}$ and external field vector $\mathbf{h}$:
\begin{equation*}
\mathbf{J}=\begin{pmatrix}
J_{11}  & J_{12} \\
J_{12}  & J_{22} \\
\end{pmatrix},
\quad\quad\quad
\mathbf{h}=\begin{pmatrix}
h_{1}  \\
h_{2} \\
\end{pmatrix}.
\end{equation*}
\noindent The existence of the thermodynamic limit of the pressure: 
\begin{equation}\label{pressione}
 p_{N}(\mathbf{J},\mathbf{h})=\frac{1}{N}\ln Z_{N}(\mathbf{J},\mathbf{h})
\end{equation}
associated to the model defined by Hamiltonian (\ref{hamiltoniana.1}) and distribution (\ref{misura.BG}) 
is shown in the paper \cite{gallo2008bipartite} exploiting an existence theorem provided for mean field model in 
\cite{bianchi2003thermodynamic}.
In this paper we deal with the exact calculation of this limit for all the values of the parameters.
 
\begin{proposition}
Consider the Hamiltonian defined in (\ref{hamiltoniana.1}) and the function: 
\begin{equation}\label{funzione.f}
f(\mu_{1},\mu_{2})=\beta g(\mu_{1},\mu_{2})-\alpha_{1}\mathscr{I}(\mu_{1})-\alpha_{2}\mathscr{I}(\mu_{2})
\end{equation}
where $g$ is given by (\ref{funzione.g}) and: 
\begin{equation}\label{entropia}
\mathscr{I}(x)=\frac{1}{2}\Big((1+x)\ln(1+x)+(1-x)\ln(1-x)\Big).
\end{equation}
Given parameters $J_{11},J_{12}, J_{22}, h_{1}, h_{2}$ and $\alpha$, the limit for large $N$ of the pressure $p_{N}(\mathbf{J},\mathbf{h})$ associated to the Hamiltonian and defined in (\ref{pressione}) is the following:
\begin{equation*}
\lim_{N\rightarrow\infty}p_{N}(\mathbf{J},\mathbf{h})=\ln 2 +\max_{(\mu_{1},\mu_{2})}f(\mu_{1},\mu_{2}).
\end{equation*}
\end{proposition}
Therefore to compute the exact value of the thermodynamic limit of the pressure $p_{N}(\mathbf{J},\mathbf{h})$, we have to 
maximize the function $f$ with respect to the variables $\mu_{1}$ and $\mu_{2}$.  Differentiating $f$ we obtain:
\begin{align*}
\frac{\partial f}{\partial \mu_{1}} (\mu_{1},\mu_{2})&=\beta(\alpha_{1}^{2}J_{11} \mu_{1}+\alpha_{1}\alpha_{2}J_{12}\mu_{2}+
\alpha_{1}h_{1})-\alpha_{1}\tanh^{-1} (\mu_{1}) \\ 
\frac{\partial f}{\partial \mu_{2}}(\mu_{1},\mu_{2}) &= \beta(\alpha_{1}\alpha_{2}J_{12}\mu_{1}+\alpha_{2}^{2}J_{22} \mu_{2}+\alpha_{2}h_{2})-\alpha_{2}\tanh^{-1} (\mu_{2})
\end{align*}
thus the mean field equations of the model are:
\begin{equation}\label{campomedio.multi}
\begin{cases}
\mu_{1} &\!\!\!\!= \tanh\Big(\beta(\alpha_{1}J_{11}\;\mu_{1}+\alpha_{2}J_{12}\;\mu_{2}+h_{1})\Big) \\
\mu_{2} &\!\!\!\!=\tanh\Big(\beta(\alpha_{1}J_{12}\;\mu_{1}+\alpha_{2}J_{22}\;\mu_{2}+h_{2})\Big).
\end{cases} 
\end{equation}

We observe that considering the symmetric case of the model, in which $\alpha=1/2$ and
$J_{11}=J_{22}$, and changing the sign of both the parameter of mutual interaction $J_{12}$ and the first component of the 
external field $h_{1}$, we obtain a system of mean field equations symmetric to the previous one with respect to the vertical axis. Therefore to analyze the critical points of the pressure functional $f$ in the symmetric case, it suffices to consider only one sign for the parameter $J_{12}$. In particular we will study the cases in which $J_{11}$ and $J_{12}$ have opposite signs in order to keep the graphical approach introduced in \cite{bidaux1967antiferromagnetisme}, useful to understand the analytical results. 

\begin{proposition}
Consider the Hamiltonian defined in (\ref{hamiltoniana.1}) whit $\alpha=1/2$, $J_{11}=J_{22}$ and $h_{1}=h_{2}=0$. Denoted with $\lambda_{M}$ and 
$\lambda_{m}$ respectively the bigger and the smaller eigenvalue of the reduced interaction matrix $\mathbf{J}$ associated to the Hamiltonian, as $J_{11}$ and $J_{12}$ have opposite signs, 
the pressure functional $f$ given by (\ref{funzione.f}) admits the following structure of critical points:
\begin{enumerate}
\item if $J_{11}>0$ and $J_{12}<0$ ($\lambda_{M}=J_{11}-J_{12}; \lambda_{m}=J_{11}+J_{12}$)
\begin{enumerate}
\item $\,0<\beta\leq\frac{2}{\lambda_{M}}\,$: the unique critical point is the origin $(0,0)$ which is a maximum point.
\item $\,\beta>\frac{2}{\lambda_{M}}\,$ for $\,\lambda_{m}<0\,$,  or $\,\frac{2}{\lambda_{M}}<\beta<\frac{2}{\lambda_{m}}\,$ for $\,\lambda_{m}>0\,$: three critical point $(0,0)$ and $\pm(\tilde{x},-\tilde{x})$  where $\tilde{x}$ is a solution of:
\begin{equation}\label{equatio.1}
\tilde{x}=\tanh\left(\frac{\beta\lambda_{M}}{2}\tilde{x}\right).
\end{equation}
The origin is an inflection point while the other two are maximum points.
\item $\,\frac{2}{\lambda_{m}}<\beta<\frac{2}{\check{\beta}\lambda_{M}}\,$ ($\lambda_{m}>0$): five critical points $(0,0)$, $\pm(\tilde{x},-\tilde{x})$ and $\pm(\hat{x},\hat{x})$ 
where $\tilde{x}$ is a solution of (\ref{equatio.1}) and $\hat{x}$ is a solution of: 
\begin{equation}\label{equatio.3}
\hat{x}=\tanh\left(\frac{\beta\lambda_{m}}{2}\hat{x}\right)
\end{equation}
and $\check{\beta}$ is defined as a solution of:
\begin{equation}\label{equatio.2}
\dfrac{\check{\beta}\tanh^{-1}\left(\sqrt{1-\check{\beta}}\right)}{\sqrt{1-\check{\beta}}}=\dfrac{\lambda_{m}}{\lambda_{M}}.
\end{equation}
The origin is a minimum point, $\pm(\tilde{x},-\tilde{x})$ are maximum points and $\pm(\hat{x},\hat{x})$ are inflection 
points.
\item $\,\beta>\frac{2}{\check{\beta}\lambda_{M}}\,$ ($\lambda_{m}>0$): nine critical points, $\pm(\tilde{x},-\tilde{x})$, $\pm(x_{1}^{*},x_{2}^{*})$, $\pm(\hat{x},\hat{x})$, $\pm(x_{2}^{*},x_{1}^{*})$, $(0,0)$ where $\tilde{x}$ is a solution of (\ref{equatio.1}), $\hat{x}$ is a solution of (\ref{equatio.3}) while for the other four points there is no analytical definition.
The origin is a minimum point, $\pm(\tilde{x},-\tilde{x})$ and $\pm(\hat{x},\hat{x})$ are maximum points. The other four can not be maximum points.
\end{enumerate}
\item if $J_{11}<0$ and $J_{12}>|J_{11}|$ ($\lambda_{M}=J_{11}+J_{12}>0; \lambda_{m}=J_{11}-J_{12}<0$)
\begin{enumerate}
\item $\,0<\beta<\frac{2}{\lambda_{M}}\,$:
the unique critical points is the origin $(0,0)$ which is a maximum point.
\item $\,\beta>\frac{2}{\lambda_{M}}\,$: three critical points $(0,0)$ and $\pm(\tilde{x},\tilde{x})$ where $\tilde{x}$ is a solution of (\ref{equatio.1}). The origin is an inflection point, 
$\pm(\tilde{x},\tilde{x})$ are maximum points.
\end{enumerate}
\item if $J_{11}<0$ and $0<J_{12}<|J_{11}|$ ($\lambda_{M}=J_{11}+J_{12}<0; \lambda_{m}=J_{11}-J_{12}<0$)\\
the unique critical points is the origin $(0,0)$ which is a maximum point.
\end{enumerate}
\end{proposition}

\begin{corollary}
Consider the Hamiltonian defined in (\ref{hamiltoniana.1}) whit $\alpha=1/2$, $J_{11}=J_{22}$ and $h_{1}=h_{2}=0$. If $J_{11}$ and $J_{12}$ have the same signs, 
the pressure functional $f$ given by (\ref{funzione.f}) admits a structure of critical points symmetric with respect to the vertical axis to those presented in Proposition 1.
\end{corollary}
The corollary follows from Proposition 2 and the properties of symmetry previously exposed.

\newpage
\begin{proposition}
Consider the Hamiltonian defined in (\ref{hamiltoniana.1}) whit $\alpha=1/2$, $J_{11}=J_{22}>0$ and $J_{12}<0$. 
\begin{enumerate}
\item As $h_{1}=h_{2}=0$ the pressure of the system is given by:
\begin{equation*}
\lim_{N\rightarrow\infty}p_{N}(\mathbf{J},\mathbf{0})=
\begin{cases}
\ln 2 + f(0,0) =\ln 2 & \text{when $ t\geq 1$} \\
\ln 2 + f(\tilde{x}(t), - \tilde{x}(t)) & \text{when $ t<1$}
\end{cases}
\end{equation*}
where $f$ is given by (\ref{funzione.f}), $\tilde{x}$ is a solution of (\ref{equatio.1}) and $t=2(\beta\lambda_{M})^{-1}$ whit $\lambda_{M}$ defined as in the previous proposition. 
\item The application of an infinitesimal field $\mathbf{h}=(h_{1},h_{2})$ with $h_{1}\neq h_{2}$ selects the state such that the scalar product between the state and the field is positive; a field with equal components retains the symmetry of the system and does not allow to select a preferred status.
\end{enumerate}
\end{proposition}
\noindent For all other possible combinations of the parameters $J_{11}$ and $J_{12}$ similar results can be obtained simply by  considering the model's symmetry, the Proposition 2 and its Corollary.

We observe that as the external field is away and the bigger eigenvalue of the reduced interaction matrix is positive, at the temperature $t=1$ the system undergoes a transition phase from zero to non-zero magnetization.
\section{Proofs}
\noindent {\bf Proof of Proposition 1}\\
We compute the exact solution of the thermodynamic limit exploiting a tail estimation on the number of configurations 
that share the same vector of the magnetization. In this way we obtain a lower and an upper bound for the partition 
function that converge to a same value as $N\rightarrow\infty$. This technique is used by Talagrand in 
\cite{talagrand2003spin} to compute the thermodynamic limit for the Curie-Weiss model.\\ 
Denoted with $\boldsymbol{\sigma}_{l}$ the configuration of the spins of the set $P_{l}$ and:
\begin{equation}\label{cardinalita}
 A_{\mu_{l}}=\card\bigg\{\boldsymbol{\sigma}_{l}\in\Omega_{N_{l}}\Big| m_{l}(\boldsymbol{\sigma})=\mu_{l}\bigg\},
\end{equation}
we can write:
\begin{equation*}
Z_{N}(\mathbf{J},\mathbf{h})=\sum_{\boldsymbol{\mu}}A_{\mu_{1}}A_{\mu_{2}}\exp\left(\beta Ng(\mu_{1},\mu_{2})\right).
\end{equation*}
\begin{lemma}\label{lemma.talagrand}
Consider the set $\Omega_{N_{l}}=\{-1,1\}^{N_{l}}$ of all possible configuration $\boldsymbol{\sigma}_{l}$. 
Let $A_{\mu_{l}}$ be a positive number defined by (\ref{cardinalita}). Then the following inequality holds:
	\begin{equation}\label{bound}
	\frac{1}{C}\frac{2^{N_{l}}}{\sqrt{N_{l}}}\exp(-N_{l}\mathscr{I}(\mu_{l}))\leq 
	A_{\mu_{l}}\leq 2^{N_{l}}\exp(-N_{l}\mathscr{I}(\mu_{l}))
	\end{equation}
	where $C$ is a constant and $\mathscr{I}$ is given by (\ref{entropia}).
\end{lemma}\vspace{0.5cm}
See appendix for the proof. This lemma allows to bound the partition function in the following way:
\begin{multline*}
\frac{1}{C}\frac{2^{N}}{\sqrt{N_{1}N_{2}}}\sum_{\boldsymbol{\mu}}\exp\Big(\beta Ng(\boldsymbol{\mu})-N_{1}\mathscr{I}(\mu_{1})-N_{2}\mathscr{I}(\mu_{2})\Big)\leq 
Z_{N}(\mathbf{J},\mathbf{h})\leq\\\leq 2^{N}\sum_{\boldsymbol{\mu}}\exp\Big(\beta Ng(\boldsymbol{\mu})-N_{1}\mathscr{I}(\mu_{1})-N_{2}\mathscr{I}(\mu_{2})\Big)
\end{multline*}
then, since a sum of positive elements is always greater than its addends and always less than the largest addend times the number of elements, we have:
\begin{equation*}
\frac{1}{C}\frac{2^{N}}{\sqrt{N_{1}N_{2}}}\exp\Big(N\max_{\boldsymbol{\mu}}f(\boldsymbol{\mu})\Big)\leq 
Z_{N}(\mathbf{J},\mathbf{h})\leq 2^{N}(N_{1}+1)(N_{2}+1)\exp\Big(N\max_{\boldsymbol{\mu}}f(\boldsymbol{\mu})\Big)
\end{equation*}
where the function $f$ is defined in (\ref{funzione.f}).
Hence for the pressure (\ref{pressione}) we have:
\begin{multline*}
\ln 2-\frac{1}{N}\bigg(\ln C+\frac{1}{2}\ln N_{1}N_{2}\bigg)+\max_{\boldsymbol{\mu}}f(\boldsymbol{\mu})
\leq p_{N}(\mathbf{J},\mathbf{h})\leq\\\leq \ln 2+\frac{1}{N}\ln((N_{1}+1)(N_{2}+1))+
\max_{\boldsymbol{\mu}}f(\boldsymbol{\mu}).
\end{multline*}
Taking the limit for $N\rightarrow\infty$ the Proposition 1 is proved.\\

\noindent{\bf Proof of Proposition 2}\\
To study the space phase of the model defined by the Hamiltonian (\ref{hamiltoniana.1}) under the hypothesis that
$\alpha=1/2$, $J_{11}=J_{22}$, $h_{1}=h_{2}=0$ and the signs of $J_{11}$ and $J_{12}$ are opposite we rescale the model's parameter with 
respect to the maximal eigenvalue $\lambda_{M}$ of the reduced interaction matrix $\mathbf{J}$ in the following way: \begin{equation}\label{riscalamento}
 a := \frac{J_{11}}{|\lambda_{M}|}; \quad b := \frac{J_{12}}{|\lambda_{M}|};  \quad
 t := \frac{2}{\beta|\lambda_{M}|}.
\end{equation}
In this way the system of mean field equations (\ref{campomedio.multi}) becomes:
\begin{equation*}
\left\{ \begin{array}{rcl} x_{1} &=& \tanh  \left[ \frac{1}{t} (a x_{1} + b  x_{2}) \right] \\ 
x_{2} &=& \tanh  \left[ \frac{1}{t} (b x_{1} + a  x_{2} ) \right]. \end{array} \right.  
\end{equation*}
By inverting the hyperbolic tangent in the two equations, we can write $x_{1}$ as a function of $x_{2}$, and
vice-versa $x_{2}$ as a function of $x_{1}$. Therefore, when $b\neq 0$ (that is the model doesn't degenerate toward two  
Curie-Weiss models) we can rewrite the equations in the following fashion:
\begin{equation*}\label{curve}
\left\{ \begin{array}{rcl} x_{2} &=& \frac{1}{b} ( - a x_{1} +  t \tanh^{-1} (x_{1}))  \\
x_{1} &=&  \frac{1}{b} ( - a x_{2} +  t \tanh^{-1} (x_{2})). \end{array} \right. 
\end{equation*}
Such a system lends itself to a graphic resolution. In fact the points that verify simultaneously the two equations are the intersections of curves $\gamma_{1}$ and $\gamma_{2}$, respectively the graphs of the functions $x_{2}(x_{1})$ and $x_{1}(x_{2})$; $\gamma_{2}$ is the symmetric curve of $\gamma_{1}$ with respect to both the bisector of the Cartesian plane. 
In particular we will plot the curve $\gamma_{1}$ by studying the function $x_2(x_1)$ and the curve $\gamma_{2}$ by symmetry. Before starting to analyze the different cases as the interaction parameters and the temperature change, we observe that $x_{2}(x_{1})$ is an odd function in the interval $(-1,1)$, it diverges as $x_{1}\rightarrow\pm 1$ depending on the sign of $b$ and its derivative are:
\begin{align}\label{derivate.1e2}
x_{2}'(x_{1}) &= \frac{1}{b} \left( \frac{t}{1-x_{1}^2} -a \right);\qquad  x_{2}'(0)= \frac{1}{b} (t-a);\\
x_{2}''(x_{1}) &= \frac{2 t}{b} \; \frac{x_{1}}{(1-x_{1}^2)^2}.\nonumber
\end{align}
We are now ready to analyze in detail the different cases.\vspace{0.3cm}\\
\noindent\textbf{1. $\mathbf{J_{11}>0}$ and $\mathbf{J_{12}<0}$}\\
The maximal eigenvalue of the reduced interaction matrix is $\lambda_{M}=J_{11}-J_{12}$, thus from (\ref{riscalamento}) we have $0\leq a< 1$,  $-1\leq b < 0$ and $a-b=1$.
In particular since b is negative, the function $x_2(x_1)$ tends to $+\infty$ as $x_{1}\rightarrow -1$ and to $-\infty$ as $x_{1}\rightarrow 1$ and the sign of its second derivative is always opposite to that of the variable.\vspace{0.3cm}\\
\noindent (a) \textit{$t>1$ (i.e. $0<\beta<\frac{2}{\lambda_{M}}$)} \\
from (\ref{derivate.1e2}) we get that the function $x_{2}(x_{1})$ is monotonic decreasing and the value of its first derivatives in the origin is minor than $-1$. Thus the curve $\gamma_{1}$ lies above the bisector of the second and fourth quadrant when $x_{1}<0$ and over when $x_{1}>0$, the symmetric curve $\gamma_{2}$  lies over the bisector when $x_{1}<0$ and above when $x_{1}>0$, therefore the unique intersection is the origin.

\begin{figure}[h!]
\centering
\includegraphics[scale=0.4]{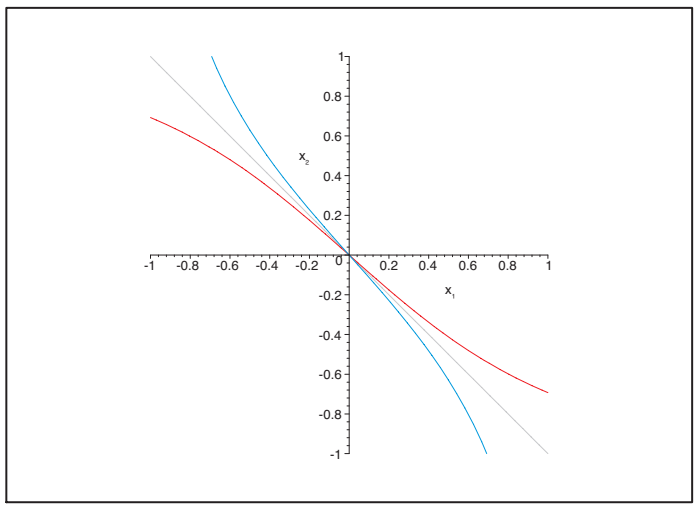}\\
\caption[\scriptsize{t=1.1, b=-0.8}]{\scriptsize{t=1.1, b=-0.8. The blue curve is $\gamma_{1}$, while the red is $\gamma_{2}$.}}
\label{fig:1}
\end{figure}
\noindent In order to understand what type of critical point is the origin, we consider the function $f$ defined in (\ref{funzione.f}), we rescaled it following (\ref{riscalamento}):
\begin{equation*}
f(x_{1},x_{2})=\frac{1}{2}\left(\frac{1}{t}\left(\frac{a}{2}\left(x_{1}^{2}+x_{2}^{2}\right)+bx_{1}x_{2}\right)-\mathscr{I}(x_{1})-\mathscr{I}(x_{2})\right),
\end{equation*}
we compute its second derivatives: 
\begin{align*}
&\frac{\partial^2 f}{\partial x_{1}^2} (x_{1},x_{2})=  \frac{1}{2t} \left(a-\frac{t}{1-x_{1}^2} \right)\\ 
&\frac{\partial^2 f}{\partial x_{2}^2} (x_{1},x_{2})=  \frac{1}{2t} \left(a-\frac{t}{1-x_{2}^2} \right)\\ 
&\frac{\partial^2 f}{\partial x_{1} \partial x_{2}} (x_{1},x_{2})= \frac{b}{2t} 
\end{align*}
and the determinant of its Hessian matrix:
\begin{equation}\label{dethessiano}
H_f (x_{1},x_{2})= \frac{1}{4t^{2}}\left(\left(a- \frac{t}{1-x_{1}^2} \right) \left(a- \frac{t}{1-x_{2}^2} \right)
 - b^2\right). 
\end{equation}
In the origin we have:
\begin{align*}
H_{f}(0,0) &=\frac{1}{4t^{2}}((a-t)^2 - b^2) = \frac{1}{4t^{2}}((b+1-t)^2 - b^2)=\nonumber\\ 
&= \frac{1}{4t^{2}}(t-1)(t-1-2 b)  \nonumber\\
\frac{\partial^2 f}{\partial x_{1}^2} (0,0) &=  \frac{1}{2t} (a-t) \nonumber
\end {align*}
\noindent Thus, since $t>1$, $0\leq a<1$ and $-1\leq b<0$ the origin is a maximum point of $f$.\vspace{0.3cm}\\
\noindent (b) \textit{$t=1$ (i.e. $\beta=\frac{2}{\lambda_{M}}$)}\\
the function $x_{2}(x_{1})$ is still monotonic decreasing and the value of its first derivatives in the origin is equal to $-1$. 
Therefore the position of the curves $\gamma_{1}$ and $\gamma_{2}$ with respect to the bisector of the second and fourth quadrant are the same of previous case. And so there is only one intersection between them in the origin.
The determinant of the Hessian matrix (\ref{dethessiano}) in this point is equal to zero. To understand what type of critical point is the origin we consider the following change of variable that diagonalizes the reduced interaction matrix: 
\begin{align}\label{diagonalizzazione}
\left\{ \begin{array}{rcl} X &=& \frac{x_1 + x_2}{2} \\ Y &=& \frac{x_2 - x_1}{2} \end{array} \right. 
\end{align}
The Taylor expansion of the function $f$ expressed in the new variable assures that such a point is a maximum point. (For the details see the Appendix).\vspace{0.3cm}\\
\noindent (c) \textit{$a\leq t<1$ (i.e. $\frac{2}{\lambda_{M}}<\beta\leq \frac{2}{J_{11}}$)}\\
the function $x_{2}(x_{1})$ is still decreasing but its derivative in the origin (\ref{derivate.1e2}) is bigger than $-1$, thus the curve $\gamma_{1}$ intersects the bisector of the second and fourth quadrant (and therefore the curve $\gamma_{2}$) in $(0,0)$ and in other two points symmetric with respect to the origin, $(-\tilde{x},\tilde{x})$ and $(\tilde{x},-\tilde{x})$, where $\tilde{x}$ is the positive solution of the equation $v(x)=0$: 
\begin{equation}\label{sol.antidiagonale}
v(x)=t \tanh^{-1}(x)-x.
\end{equation}
\begin{figure}[h!]
\centering
\includegraphics[scale=0.4]{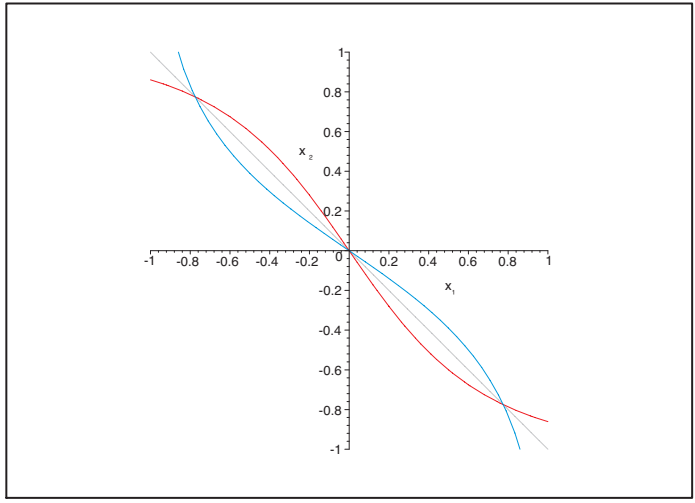}\\
\caption[\scriptsize{t=0.75, b=-0.8}]{\scriptsize{t=0.75, b=-0.8. The blue curve is $\gamma_{1}$, while the red is $\gamma_{2}$.}}
\label{fig:2a}
\end{figure}\\
Since the function, $v$ given by (\ref{sol.antidiagonale}), intersects the horizontal axis in the origin, reaches its minimum in $x=\sqrt{1-t}$ and goes to infinity for large $x$, it follows:
\begin{equation}\label{utile}
\tilde{x}>\sqrt{1-t}.
\end{equation}
\noindent This inequality turns to be useful to evaluate the nature of the critical points $\pm(\tilde{x},-\tilde{x})$. In these points the determinant of the Hessian matrix of the function $f$ (\ref{dethessiano}) is: 
\begin{equation}\label{dettilde}
H_{f}(-\tilde{x},\tilde{x})=H_{f}(\tilde{x},-\tilde{x}) =\frac{1}{4t^2} \left( \left(a-\frac{t}{1 - \tilde{x}^2}  \right)^2 - b^2\right) .
\end{equation}
Since in the considered range of temperature the function on the r.h.s of (\ref{dettilde}) is monotonic increasing for positive $x$, by the inequality (\ref{utile}) we have:
\begin{align*}
H_{f}(-\tilde{x},\tilde{x})=H_{f}(\tilde{x},-\tilde{x}) & > \frac{1}{4t^2} \left(\left(a- \frac{t}{1 - (1-t)}  \right)^2 - b^2 \right)=\\
&=\frac{1}{4t^2}\left( (a-1)^2 - b^2\right) \; = \; 0.
\end{align*}
On the other hand the first term of the Hessian matrix in $\pm(\tilde{x},-\tilde{x})$ is a decreasing function and thus:
\begin{align*}
\frac{\partial^2 f}{\partial x^2}(-\tilde{x},\tilde{x}) = \frac{\partial^2 f}{\partial x^2}(\tilde{x},-\tilde{x}) &=  \frac{1}{2t} \left( a-\frac{t}{1 - \tilde{x}^2} \right)<\\ 
 &<  \frac{1}{2t} (a-1) \, < \, 0.
\end{align*} 
Then $\pm(\tilde{x},-\tilde{x})$ are maximum points. On the contrary the origin is an inflection point because the determinant of the Hessian matrix (\ref{dethessiano}) computed in such a point is negative for the considered range of temperature.\vspace{0.3cm}\\
\noindent (d) \textit{ $2a-1<t<a$ with $2a-1>0$ or $0<t<a$ (i.e. $\lambda_{m}>0$ and $\frac{2}{J_{11}}<\beta<\frac{2}{\lambda_{m}}$ or\linebreak\indent $\;\;\lambda_{m}<0$ and $\beta>\frac{2}{J_{11}}$)}\\
\noindent The function $x_{2}(x_{1})$ has a minimum point in $x_{m}=-\sqrt{1-t/a}$ and a maximum point in $x_{M}=\sqrt{1-t/a}$.
Therefore the curves $\gamma_{1}$ and $\gamma_{2}$ intersect each other in at least three points: the origin and $\pm(\tilde{x},-\tilde{x})$, where $\tilde{x}$ is again the positive solution of $v(x)=0$, with $v$ defined in (\ref{sol.antidiagonale}). However, since in the origin the first derivative of $x_{2}(x_{1})$ is positive and less than one, $\gamma_{1}$ lies above the bisector of the first and third quadrant when $x_{1}<0$ and over when $x_{1}>0$. Instead $\gamma_{2}$ lies over the bisector when $x_{1}<0$ and above when $x_{1}>0$. This scenario excludes the possibility of additional intersections between the two curves.
\begin{figure}[h!]
\centering
\includegraphics[scale=0.4]{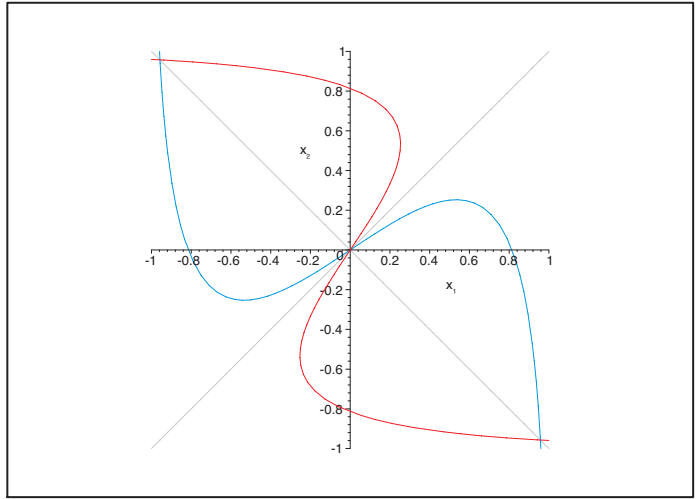}\\
\caption[\scriptsize{t=0.5, b=-0.3}]{\scriptsize{t=0.5, b=-0.3. The blue curve is $\gamma_{1}$, while the red is $\gamma_{2}$.}}
\label{fig:2b}
\end{figure}\\
 Again $\pm(\tilde{x},-\tilde{x})$ are maximum points and the origin is an inflection point.\vspace{0.3cm}\\
\noindent (e) \textit{$t=2a-1$ with $2a-1>0$ (i.e. $\lambda_{m}>0$ and $\beta=\frac{2}{J_{11}}$)}\\
in this case the only difference from the previous one is the value of the first derivative of the function $x_{2}(x_{1})$ at the origin, which now is exactly equal to 1. This does not imply any qualitative change in the relative positions of curves $\gamma_{1}$ and $\gamma_{2}$, that continue to intersect in the points $(-\tilde{x},\tilde{x})$, $(0,0)$ and $(\tilde{x},-\tilde{x})$.
The analysis of the Hessian matrix of the function $f$ assures that $\pm(\tilde{x},-\tilde{x})$ are still maximum points but does not give information about the nature of the origin. Considering again the change of variable (\ref{diagonalizzazione}) we can conclude that the origin is an inflection point. (See the Appendix).\vspace{0.3cm}\\
\noindent (f) \textit{$\check{t}\leq t<2a-1$ where $\check{t}$ is a solution of $\;\check{t}\tanh^{-1}(\sqrt{1-\check{t}})-\sqrt{1-\check{t}}=2b\sqrt{1-\check{t}}$ \linebreak\indent (i.e. $\,\lambda_{m}>0\,$ and $\,\frac{2}{\lambda_{m}}<\beta<\frac{2}{\check{\beta}\lambda_{M}}\,$ where $\check{\beta}$ is a solution of (\ref{equatio.2})}\\
the function $x_{2}(x_{1})$ has the same intervals of monotony of the previous cases but the value of its first derivatives in the origin is bigger than $1$. This implies that it intersect the bisector of the first and third quadrant in two other points beyond the origin, $(-\hat{x},-\hat{x})$ and $(\hat{x},\hat{x})$, where $\hat{x}$ is the positive solution of the equation $v(x)=2bx$, where $v$ is the function defined in (\ref{sol.antidiagonale}). 
Thus the curves $\gamma_{1}$ and $\gamma_{2}$ intersect each other in at least five points: $(0,0)$, $\pm(\tilde{x},-\tilde{x})$, $\pm(\hat{x},\hat{x})$. The graphic representation presented in \cite{bidaux1967antiferromagnetisme} excludes the possibility of further solutions that will appear as $t<\check{t}$.
\begin{figure}[h!]
\centering
\includegraphics[scale=0.4]{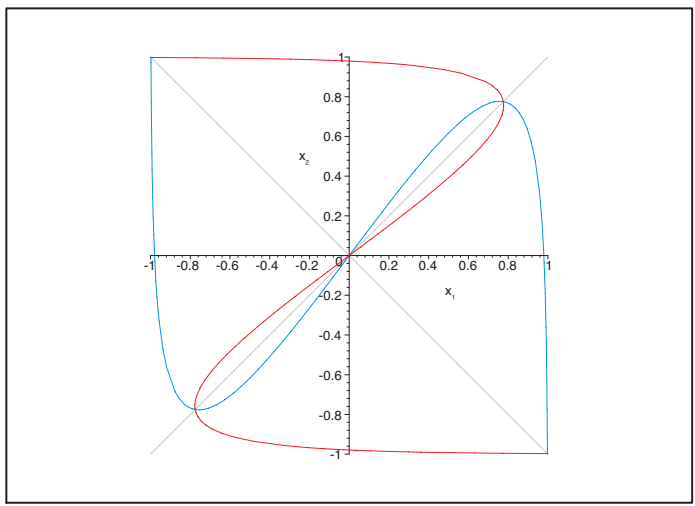}\\
\caption[\scriptsize{t=0.3, b=-0.3}]{\scriptsize{t=0.3, b=-0.3. The blue curve is $\gamma_{1}$, while the red is $\gamma_{2}$.}}
\label{fig:3}
\end{figure}\\
\noindent The computations made on the Hessian matrix assures that the origin is a minimum point and $\pm(\tilde{x},-\tilde{x})$ are maximum points.   
The new solutions $\pm(\hat{x},\hat{x})$ are inflection points because in these points the determinant of the Hessian matrix of the function $f$ can be expressed in the following way:
\begin{align*}
H_{f}(-\hat{x},-\hat{x})=H_{f}(\hat{x},\hat{x}) &=\frac{1}{4t^2} \left( \left(a-\frac{t}{1 - \hat{x}^2}  \right)^2 - b^2\right)= \nonumber\\
& = \frac{b^{2}}{4t^2} \left(\left(x_{2}'(\hat{x}) \right)^2 - 1\right)
\end{align*}
where $-1\leq\left(x_{2}'(\hat{x}) \right)<1$ in the considered range of temperature. To prove this statement we consider the inequality:
\begin{equation*}
\hat{x}>\sqrt{1-\frac{t}{1+2b}}
\end{equation*}
true for the solutions of the equation $v(x)=2bx$ for the same reason for which the inequality (\ref{utile}) is true for the solution of $v(x)=0$. Since:
\begin{equation*}
x_{2}'\left(\sqrt{1-\frac{t}{1+2b}}\right)=1
\end{equation*}
and the function $x_{2}'(x_{1})$ is monotonic decreasing for positive values ​​of the abscissa, we have that $x_{2}'(\hat{x})<1$. In particular the value of this derivative, initially positive, decreases as the temperature $t$ decreases till the values $-1$ that it reaches for $t=\check{t}$. At this temperature the determinant of the Hessian matrix computed in $\pm(\hat{x},\hat{x})$ is equal to zero. Nevertheless, without need to resort to the diagonalization of the reduced interaction matrix, we can say that these points, equal to $\pm(\sqrt{1-t},\sqrt{1-t})$, are inflection points. In fact: 
\begin{equation*}
\frac{\partial^3 f}{\partial x_{1}^3}(\pm\sqrt{1-t},\pm\sqrt{1-t})=\frac{\partial^3 f}{\partial x_{2}^3}(\pm\sqrt{1-t},\pm\sqrt{1-t})=-\dfrac{\sqrt{1-\check{t}}}{\check{t}^{2}}\neq 0
\end{equation*} 
and the other partial derivatives of the third order of the function $f$ are equal to zero.\vspace{0.3cm}\\
\noindent (g) \textit{$0<t<\check{t}$ (i.e. $\,\lambda_{m}>0\,$ and $\,\beta>\frac{2}{\check{\beta}\lambda_{M}}\,$)}\\
the graphical representation presented in \cite{bidaux1967antiferromagnetisme} shows that as the temperature decreases below the value $\check{t}$ a further solution $(x_{1}^{*},x_{2}^{*})$ of the mean field equations appear, with $x_{2}^{*}>x_{1}^{*}>0$; the other three solutions are generated by the symmetry with respect to both the bisector. In these points it is not always possible to detect the sign of the determinant of the Hessian matrix.
In any case, since they are linked to a maximum point by a monotonic piece of one of the two curves $\gamma_{1}$ and $\gamma_{2}$, we can exclude that they are in turn maximum points. (For further details see \cite{gallo2008bipartite}). When $t$ is small enough it can be shown analytically that they are inflection points.\\
\begin{figure}[h!]
\centering
\includegraphics[scale=0.4]{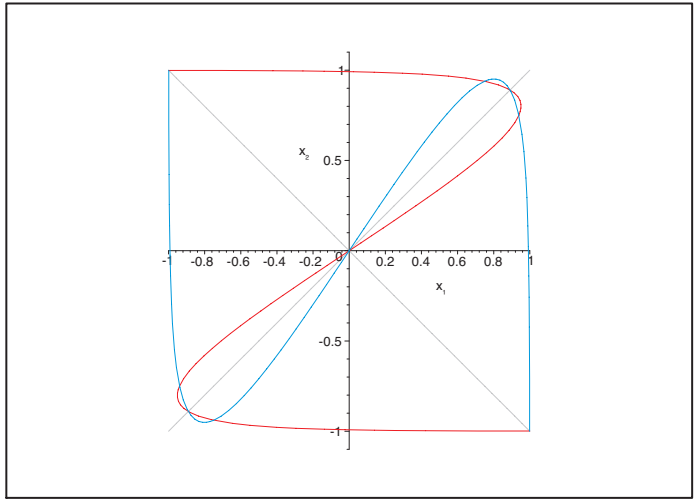}\\
\caption[\scriptsize{t=0.25, b=-0.3}]{\scriptsize{t=0.25, b=-0.3. The blue curve is $\gamma_{1}$, while the red is $\gamma_{2}$.}}
\label{fig:4}
\end{figure}\\
 As in the previous case the origin is a minimum point and $\pm(\tilde{x},-\tilde{x})$ are maximum points; on the contrary $\pm(\hat{x},\hat{x})$ became maximum points. In fact since $x_{2}'(\hat{x})<-1$ as $t<\check{t}$, the determinant of the Hessian matrix of $f$ in these points is positive and: 
\begin{equation*}
\frac{\partial^2 f}{\partial x_{1}^2}(\hat{x},\hat{x})=\frac{\partial^2 f}{\partial x_{1}^2}(-\hat{x},-\hat{x})=-\dfrac{b}{2t}x_{2}'(\hat{x})<0.
\end{equation*}\vspace{0.3cm}\\
\noindent \textbf{2. $\mathbf{J_{11}<0}$ and $\mathbf{J_{12}>|J_{11}|}$}\\
The maximal eigenvalue of the reduced interaction matrix is $\lambda_{M}=J_{11}+J_{12}$, thus from (\ref{riscalamento}) we have $a<0$,  $ b>1$ and $a+b=1$.
In particular since b is positive, the function $x_2(x_1)$ tends to $-\infty$ as $x_{1}\rightarrow -1$ and to $+\infty$ as $x_{1}\rightarrow 1$ and the sign of the second derivative of $x_2(x_1)$ is always the same of the variable. Moreover the signs of $a$ and $b$ assure that the function $x_{2}(x_{1})$ is always monotonic increasing. Therefore we can distinguish only two scenarios:\vspace{0.3cm}\\
\noindent (a) \textit{$t\geq 1$ (i.e. $0<\beta\leq \frac{2}{\lambda_{M}}$)}\\
the value of the first derivative of $x_{2}(x_{1})$ computed in the origin is major or equal to $1$. Thus $\gamma_{1}$ lies over the bisector of the first and third quadrant when $x_{1}<0$ and above when $x_{1}>0$. Therefore since $\gamma_{2}$ is symmetric to $\gamma_{1}$ with respect to this bisector the curves intersect each other once in the origin. Analyzing the Hessian matrix of the function $f$ in this point for $t>1$ we have: 
\begin{align*}
H_{f}(0,0) &= \frac{1}{4t^{2}}(t-1)(t-1+2 b) \, > \, 0 \\
\frac{\partial^2 f}{\partial x_{1}^2} (0,0) &=  \frac{1}{2t} (a-t) \; < \; 0.
\end{align*}
Thus $(0,0)$ is a maximum point.\\
As $t=1$ in the origin, the determinant of the Hessian matrix of the function $f$ is equal to zero, however considering the change of variable (\ref{diagonalizzazione}) and the Taylor expansion of the function $f$ we get that such a point is a maximum point. (See the Appendix).\vspace{0.3cm}\\
\noindent (b) \textit{$0<t<1$ (i.e. $\beta>\frac{2}{\lambda_{M}}$)}\\
the value of the first derivative of $x_{2}(x_{1})$ in the origin is less than $1$ and then the curves $\gamma_{1}$ and $\gamma_{2}$ intersect each other in three point: $(0,0)$ and other two points symmetric with respect to the origin $(-\tilde{x},-\tilde{x})$, $(\tilde{x},\tilde{x})$ where $\tilde{x}$ is the positive solution of the equation $v(x)=0$.\\
\noindent By inequalities (\ref{utile}) we have:
\begin{align*}
H_{f}(-\tilde{x},-\tilde{x})=H_{f}(\tilde{x},\tilde{x}) & > \frac{1}{4t^2} \left(\left( \frac{t}{1 - (1-t)} - a \right)^2 - b^2 \right)=\\
&=\frac{1}{4t^2}\left( (1-a)^2 - b^2\right) \; = \; 0\\
\frac{\partial^2 f}{\partial x^2}(-\tilde{x},-\tilde{x})= \frac{\partial^2 f}{\partial x^2}(\tilde{x},\tilde{x}) &=  \frac{1}{2t} \left( a-\frac{t}{1 - \tilde{x}^2}  \right)= \\ 
&=  \frac{1}{2t} (a-1) \, < \, 0.
\end{align*}
Therefore $\pm(\tilde{x},-\tilde{x})$ are maximum points. On the other hand since:
\begin{equation*}
H_{f}(0,0)= \frac{1}{4t^{2}}(t-1)(t-1+2 b)\;<0
\end{equation*}
the origin is an inflection point.\vspace{0.3cm}\newpage
\noindent \textbf{3. $\mathbf{J_{11}<0}$ and $\mathbf{0<J_{12}<|J_{11}|}$}\\
The maximal eigenvalue of the reduced interaction matrix, $\lambda_{M}=J_{11}+J_{12}$, is negative. Therefore by (\ref{riscalamento}) we have $a<-1$,  $ b>0$ and $a+b=-1$. Since the signs of parameters $a$ and $b$ are the same of the previous case, the comments made about the limit behavior, the monotony and the concavity of the function $x_{2}(x_{1})$ are still valid. Furthermore, since $a>-1$ the value of the first derivative of $x_{2}(x_{1})$ computed in the origin is bigger than $1$ for all $t>0$. As we have seen before this imply that curves $\gamma_{1}$ and $\gamma_{2}$ intersect each other only in the origin. Since:
\begin{align*}
H_{f}(0,0) &= \frac{1}{4t^{2}}(t+1)(t+2 b+1) \, > \, 0 \\ 
\frac{\partial^2 f}{\partial x_{1}^2} (0,0) &=  \frac{1}{2t} (a-t) \; < \; 0,
\end{align*}
such a point is a maximum point.\\

\noindent{\bf Proof of Proposition 3}
\begin{enumerate}
\item 
If $2a-1<0$ the statement follows directly from Proposition 2. If $2a-1>0$ the same is true only for $t>\check{t}$.
For $0<t<\check{t}$ the function  $f$ admits four maximum points: $\pm(\tilde{x},-\tilde{x})$ and $\pm(\hat{x},\hat{x})$, therefore to find what is the absolute maximum between them we have to compare the values of the function in these points. For this purpose it is more convenient to consider the function $P(x_1, x_2)=tf(x_1, x_2)$:
\begin{equation*}
P(x_1, x_2) = \frac{1}{4} a (x_1^2 + x_2^2) + \frac{1}{2} b x_1 x_2 - \frac{t}{2} \left( \mathscr{I}(x_1) + \mathscr{I}(x_2) \right).
\end{equation*}
Since the function $P$ is pair we choose to compute the value it assumes in the points with positive abscissa, $(\tilde{x},-\tilde{x})$ and $(\hat{x},\hat{x})$ where $\tilde{x}$ and $\hat{x}$ are positive:
\begin{align*}
P(\tilde{x}, -\tilde{x}) &= \frac{1}{2} a \tilde{x}^2 - \frac{1}{2}  b \tilde{x}^2 - t \mathscr{I}(\tilde{x})  =  \frac{1}{2} \tilde{x}^{2} - t \mathscr{I}(\tilde{x}) \\
P(\hat{x},\hat{x}) &= \frac{1}{2} a \hat{x}^2 + \frac{1}{2}  b \hat{x}^2 - t \mathscr{I}(\hat{x})  = \frac{1}{2} (a+b) \hat{x}^{2} - t \mathscr{I}(\hat{x}). 
\end{align*}
We observe that in the points $(\tilde{x},-\tilde{x})$ and $(\hat{x},\hat{x})$, the first and the second derivatives with respect to the temperature of the function $P$ can be write:
\begin{equation*}
\frac{dP(M(t))}{dt}=-\mathscr{I}(x_{M}(t));\qquad \frac{d^{2}P((M(t))}{dt^{2}} =-\mathscr{I}(x_{M}(t))\,x_{M}'(t)
\end{equation*}
where $M(t)$ denotes in a generic way one of the two maximum points and $x_{M}(t)$ is its abscissa. Since the function $\mathscr{I}$ is pair, convex and monotonic increasing for positive abscissa, and $x_{M}(t)$ is decreasing, the function $P$ is decreasing ad concave along the maximun points.
As the temperature tends to zero both $\tilde{x}$ and $\hat{x}$ tend to one, therefore: 
\begin{equation*}
\lim_{t\rightarrow 0}P(\tilde{x}, -\tilde{x}) =\frac{1}{2},\quad\qquad
\lim_{t\rightarrow 0}P(\hat{x},\hat{x}) = \frac{2a-1}{2}.
\end{equation*}
Since $0<a<1$ we obtain that as $t\rightarrow 0$ the pressure reaches the higher value in $(\tilde{x}, -\tilde{x})$. 
Also as $t=\check{t}$ it is easy to show that:
\begin{equation*}
P(\hat{x}, \hat{x})<P(\tilde{x}, -\tilde{x}).
\end{equation*}
Thus, since $\tilde{x}(t)>\hat{x}(t)$ as $0<t<\check{t}$ we conclude that: 
\begin{equation*}
P(\tilde{x},-\tilde{x},t)>P(\hat{x},\hat{x},t)\quad\quad \text{for } 0<t<\check{t}.
\end{equation*}
\item 
If we apply an infinitesimal field $\mathbf{h}=(h_{1},h_{2})$ to the considered system, the maximum points of the function $f$ remain around the points that were maximum points when the field was away. We are interested in showing the field's ability to select a preferred state, thus we omit the case $t\geq 1$ in which there is only one possible state. On the contrary for other temperature values the function $f$ admits at least two maximum points. The gap between the values that the function $P$ assumes in the points $\pm(\tilde{x},-\tilde{x})$ and $\pm(\hat{x},\hat{x})$ in absence of the field assures that when the field is infinitesimal the maximum points close to $\pm(\hat{x},\hat{x})$ cannot be the effective state of the system. 
In order to understand which of the two remaining points is the preferred we distinguish three cases.\\
If the field components are equal, $\mathbf{h}=(h_{1},h_{1})$, the mean field equations and their solutions retain the symmetry with respect to the bisector of the first and third quadrant that they presented as the field was away. Therefore for each temperature's value the pressure computed in the two global maximum points is the same and the field $\mathbf{h}$ is not able to select a preferred state between them.\\
If the field components are opposite, $\mathbf{h}=(h_{1},-h_{1})$, the system retain the symmetry with respect to the bisector of the second and fourth quadrant, thus we have two solutions $s^{(1)}$ and $s^{(2)}$ on this bisector; their first coordinates are respectively the positive and negative solution of the mean field equation:
\begin{equation*}
x_{1} =\tanh \left( \frac{1}{t} \left( x_{1} +h_{1} \right) \right) .
\end{equation*}
Therefore between $s^{(1)}$ and $s^{(2)}$, that are on opposite side with respect to the origin, the field selects the point whose direction is identical to its own.\\
\noindent At least we consider the case in which the field components are generic, that is $\mathbf{h}=(h_{1},h_{2})$ with $|h_{1}|\neq |h_{2}|$.
Decomposed the field along the eigenvector of the reduced interaction matrix $\mathbf{J}$:
\begin{equation*}
(h_1, h_2) = (h', -h')+(h'', h'')
\end{equation*}
we apply the field components in two steps. Firstly considering only $(h', -h')$, as we have seen above, we obtain two solutions $s^{(1)}=(\tilde{x}^{(1)},-\tilde{x}^{(1)})$ and $s^{(2)}$ respectively close to $(\tilde{x},-\tilde{x})$ and $(-\tilde{x},\tilde{x})$. The subsequent  application of $(h'', h'')$ causes an infinitesimal shift of $s^{(1)}$ and $s^{(2)}$ in this direction. We look for the coordinates of $s^{(1)}$ in the form: 
\begin{align*}
x_1^{(1)} (t) &= \tilde{x}^{(1)} (t) + x''_1 (t) \\
x_2^{(1)} (t) &= - \tilde{x}^{(1)} (t) + x''_1 (t)
\end{align*}
where $x'_1$ is the infinitesimal shift in the diagonal direction. From the mean field equation we obtain:
\begin{equation*}
\left\{ \begin{array}{rcl} \tilde{x}^{(1)} + x''_1 &=& \tanh \left[ \frac{1}{t} \left( \tilde{x}^{(1)} + (a+b) x''_1 +
                                                       h'' \right) \right] \\
                          - \tilde{x}^{(1)} + x''_1 &=& \tanh \left[ \frac{1}{t} \left( - \tilde{x}^{(1)} + (a+b) x''_1 +
                                                   h'' \right) \right]. \end{array} \right. 
\end{equation*}
Developing the hyperbolic tangent and neglecting the infinitesimal terms of second order of the type $x''h''$ we obtain two equations both equal to the following:
\begin{equation*}
x''_1 = \left(1 - (\tilde{x}^{(1)})^2 \right) \tanh\left[ \frac{a+b}{t} x''_1 + \frac{h''}{t} \right]
\end{equation*}
whose principal solution is those that has the same sign of $h''$. 
Since the same argument applies to the second solution we conclude that the field $\mathbf{h}=(h_{1},h_{2})$ with $|h_{1}|\neq |h_{2}|$ moves $\pm(\tilde{x},-\tilde{x})$ in two solutions which remain on opposite half-plane with respect to the bisector of the first and third quadrant. Furthermore the field ability to broke the symmetry of the function $f$ ensures that the absolute maximum point is the one whose scalar product with the field is positive.
\end{enumerate}
\section{Comments}
In this paper we have deeply investigated the behaviour of the symmetric bipartite 
mean-field model as the external field is away or small. In particular, by studying the nature of the critical points of the pressure functional we have highlighted for which values of parameters the model undergoes a phase transition. In addition to being interesting in themselves, the obtained results can provide good tools to describe more complex models, like the bipartite non-symmetric mean-field model, already used in the statistical approach to socio-economical
sciences \cite{contucci2007modeling}, or the multipartite symmetric mean-field model, that seems to have a
very interesting behaviour with non trivial phase transitions. These models will be subject of further investigation of the authors. By analyzing them is expected to relate the critical and non-critical phases to different limiting behaviours of the sums of the spins of such models \cite{fedele2011scaling}.

\section*{Appendix}
\appendix
\section{Proof of lemma \ref{lemma.talagrand}}
As $m_{l}(\boldsymbol{\sigma})=\mu_{l}$, the configuration $\boldsymbol{\sigma}_{l}$ contains $N_{l}(1+\mu_{l})/2$ times the value $1$ and $N_{l}(1-\mu_{l})/2$ times the value $-1$, thus we have:
\begin{equation*}
 A_{\mu_{l}}=
\begin{pmatrix}
 N_{l}\\
\frac{N_{l}(1+\mu_{l})}{2}
\end{pmatrix}
\end{equation*}
Using Stirling's formula, $n!\sim n^{n}e^{-n}\sqrt{2\pi n}$, we get:
\begin{align*}
 A_{\mu_{l}}&\geq\sqrt{\frac{2}{\pi}}\dfrac{1}{\sqrt{N_{l}(1-\mu_{l}^{2})}}\dfrac{N_{l}^{N_{l}}}{\Big(\frac{N_{l}(1+\mu_{l})}{2}\Big)^{N_{l}(1+\mu_{l})/2}\Big(\frac{N_{l}(1-\mu_{l})}{2}\Big)^{N_{l}(1-\mu_{l})/2}}\geq\nonumber\\
&\geq\frac{1}{C}\frac{2^{N_{l}}}{\sqrt{N_{l}}}\dfrac{1}{(1+\mu_{l})^{N_{l}(1+\mu_{l})/2}(1-\mu_{l})^{N_{l}(1-\mu_{l})/2}}=\nonumber\\
&=\frac{1}{C}\frac{2^{N_{l}}}{\sqrt{N_{l}}}\exp(-N_{l}\mathscr{I}(\mu_{l}))
\end{align*}
In this way we obtain a lower bound of $A_{\mu_{l}}$.\\ 
To obtain an upper bound for $A_{\mu_{l}}$ we suppose the spins $\sigma_{i}$ are independent Bernoulli random variable. In this case $P(\sigma_{i}=1)=P(\sigma_{i}=-1)=1/2$ and all configurations $\boldsymbol{\sigma}_{l}$ have same probability, hence:
\begin{equation*}
A_{\mu_{l}}=2^{N_{l}}P\bigg\{m_{l}(\boldsymbol{\sigma})=\mu_{l}\bigg\}\leq 2^{N_{l}}P\bigg\{m_{l}(\boldsymbol{\sigma})\geq \mu_{l}\bigg\}
\end{equation*}
where by definition of the magnetization:
\begin{equation*}
 P\bigg\{m_{l}(\boldsymbol{\sigma})\geq \mu_{l}\bigg\}=P\bigg\{\sum_{i\in P_{l}}\sigma_{i}\geq \mu_{l} N_{l}\bigg\}.
\end{equation*}
Take $\lambda>0$ by Chebyshev's inequality we can bound the above probability:
\begin{align*}
P\bigg\{\sum_{i\in P_{l}}\sigma_{i}\geq \mu_{l} N_{l}\bigg\}&\leq e^{-\lambda\mu_{l} N_{l}}\prod_{i=1}^{N_{l}}E[\exp(\lambda\sigma_{i})]=\nonumber\\
&=\exp(N_{l}(-\lambda\mu_{l} +\ln\cosh\lambda))\leq\nonumber\\
&\leq\min_{\lambda}\{\exp(N_{l}(-\lambda\mu_{l}+\ln\cosh\lambda))\}
\end{align*}
where $E$ denotes the expectation value.\\ 
If $|\mu_{l}|<1$, the previous exponent 
is minimized for:
\begin{equation}\label{talagrand.passaggio.5}
 \lambda=\tanh^{-1}(\mu_{l})=\frac{1}{2}\ln\bigg(\dfrac{1+\mu_{l}}{1-\mu_{l}}\bigg).
\end{equation}
Since $1/(\cosh^{2}y)=1-\tanh^{2}y$ the following equality holds:
\begin{equation}\label{talagrand.passaggio.6}
 \ln\cosh\lambda=-\frac{1}{2}\ln(1-\mu_{l}^{2}).
\end{equation}
Thus by (\ref{talagrand.passaggio.5}) and (\ref{talagrand.passaggio.6}): 
\begin{equation*}
\min_{\lambda}\{\exp(N_{l}(-\lambda\mu_{l}+\ln\cosh\lambda))\}=\exp(-N_{l}\mathscr{I}(\mu_{l})).
\end{equation*}
Hence the upper bound for $A_{\mu_{l}}$ is:
\begin{equation*}
A_{\mu_{l}}\leq 2^{N_{l}}\exp(-N_{l}\mathscr{I}(\mu_{l})).
\end{equation*}

\section{Diagonalization}
In the proof of proposition 1 we have seen that in three cases the determinant of the Hessian matrix of the function $f$, given by (\ref{funzione.f}), is equal to zero when computed in the critical point $(0,0)$. In particular this happen as $J_{11}>0$ and $J_{12}<0$ for $t=1$ and $t=2b+1$ and as $J_{11}<0$ and $J_{12}>|J_{11}|$ for $t=1$. Thus, in order to determinate the nature of the origin we consider the change of variable that diagonalize the reduced interaction matrix $\mathbf{J}$: 
 \begin{align*}
\left\{ \begin{array}{rcl} X &=& \frac{x_1 + x_2}{2} \\ Y &=& \frac{x_2 - x_1}{2} \end{array} \right. \quad
\Longrightarrow\quad \left\{ \begin{array}{rcl} x_1 &=& X-Y  \\ x_2 &=& X + Y \end{array} \right. 
\end{align*}
we apply it to the function $f$:
\begin{equation*}
f(X,Y) = \frac{1}{2}\left(\frac{1}{t}\left((a+b)X^{2}+(a-b)Y^{2}\right)-\mathscr{I}(X-Y)-\mathscr{I}(X+Y)\right)
\end{equation*}
and we compute the partial derivatives of $f$ with respect to the new variables. Obviously the derivatives of the first order:   
\begin{align*}
\frac{\partial f}{\partial X} (X,Y) &= \frac{a+b}{t} X - \frac{1}{2} \left( \tanh^{-1} (X-Y) + \tanh^{-1} (X+Y) \right) \\
\frac{\partial f}{\partial Y} (X,Y) &= \frac{a-b}{t} Y - \frac{1}{2} \left( \tanh^{-1} (X+Y) - \tanh^{-1} (X-Y) \right)
\end{align*}
are equal to zero in the origin. The derivatives of the second order are:
\begin{align*}
\frac{\partial^2 f}{\partial X^2} (X,Y) &= \frac{a+b}{t} - \frac{1}{2} \left( \frac{1}{1-(X+Y)^2} + \frac{1}{1-(X-Y)^2} \right) \\
\frac{\partial^2 f}{\partial Y^2} (X,Y) &= \frac{a-b}{t} - \frac{1}{2} \left( \frac{1}{1-(X+Y)^2} + \frac{1}{1-(Y-X)^2} \right) \\
\frac{\partial^2 f}{\partial X \partial Y} (X,Y) &= - \frac{1}{2} \left( \frac{1}{1-(X+Y)^2} - \frac{1}{1-(X-Y)^2} \right).
\end{align*}
In particular in $(0,0)$ we have: 
\begin{align*}
&\frac{\partial^2 f}{\partial X^2} (0,0) = \frac{a+b}{t}-1\\
&\frac{\partial^2 f}{\partial Y^2} (0,0)= \frac{a-b}{t}-1\\
&\frac{\partial^2 f}{\partial X\partial Y} (0,0)=0
\end{align*}
Thus, since $a-b=1$ as $J_{11}>0$ and $J_{12}<0$ and $a+b=1$ as $J_{11}<0$ and $J_{12}>|J_{11}|$, in all the three cases the determinant of the Hessian matrix is equal to zero. This is the reason why we need to compute partial derivatives of order higher than the second. Those of the third order are:
\begin{align*}
\frac{\partial^3 f}{\partial X^3} (X,Y)=\frac{\partial^3 f}{\partial X \partial Y^2} (X,Y)&= -\frac{1}{2} \left(\frac{2(X+Y)}{(1-(X+Y)^2)^2} + \frac{2(X-Y)}{(1-(X-Y)^2)^2} \right) \\
\frac{\partial^3 f}{\partial Y^3} (X,Y)=\frac{\partial^3 f}{\partial X^2 \partial Y} (X,Y) &= -\frac{1}{2} \left(\frac{2(X+Y)}{(1-(X+Y)^2)^2} + \frac{2(Y-X)}{(1-(X-Y)^2)^2} \right). 
\end{align*}
It is easy to check that these derivatives computed in $(0,0)$ are equal to zero. At least the partial derivative of the fourth order are:
\begin{align*}
&\frac{\partial^4 f}{\partial X^4}(X,Y) =\frac{\partial^4 f}{\partial Y^4}(X,Y) =\frac{\partial^4 f}{\partial X^2 \partial Y^2}(X,Y)=-\left(\frac{1}{(1-(X+Y)^2)^2} \right. + \\
&\qquad\qquad\qquad\qquad\left. +\frac{1}{(1-(X-Y)^2)^2} + \frac{4(X+Y)^2}{(1-(X+Y)^2)^3} + \frac{4(X-Y)^2}{(1-(X-Y)^2)^3} \right) \\\\
&\frac{\partial^4 f}{\partial X^3 \partial Y}(X,Y) =\frac{\partial^4 f}{\partial X \partial Y^3} (X,Y)= -\left(\frac{1}{(1-(X+Y)^2)^2} \right. +\\
&\qquad\qquad\qquad\qquad \left. - \frac{1}{(1-(X-Y)^2)^2}+ \frac{4(X+Y)^2}{(1-(X+Y)^2)^3} - \frac{4(X-Y)^2}{(1-(X-Y)^2)^3} \right). 
\end{align*}
In particular in $(0,0)$ these derivatives become:
\begin{align*}
\frac{\partial^4 f}{\partial X^4} (0,0) = \frac{\partial^4 f}{\partial Y^4} (0,0)&=\frac{\partial^4 f}{\partial X^2 \partial Y^2} (0,0) =-2\\
\frac{\partial^4 f}{\partial X^3 \partial Y} &= \frac{\partial^4 f}{\partial X \partial Y^3}  =  0.
\end{align*}
Now we can approximate the function $f$ in a neighborhood of the origin with its Taylor expansion till the fourth order. In particular as $J_{11}>0$ and $J_{12}<0$ and $t=1$ we have:
\begin{equation*}
 f(X,Y)=2bX^{2}-\frac{1}{12}(X+Y)^{2}+o(X^{2}+Y^{2}).
\end{equation*}
Since for these values of the parameters $b$ is negative, this polynomial is negative. Therefore the origin is a maximum point. As $J_{11}>0$ and $J_{12}<0$ and $t=2b+1$ the Taylor approximation of $f$ is:
\begin{equation*}
 f(X,Y)=-2bY^{2}-\frac{1}{12}(X+Y)^{2}+o(X^{2}+Y^{2}).
\end{equation*}
It is easy to check that this polynomial is negative when computed in the point of the horizontal axis and positive in the point of the vertical axis. Thus the origin is an inflection point. At least as $J_{11}<0$ and $J_{12}>|J_{11}|$ and $t=1$ we have:
\begin{equation*}
 f(X,Y)=-2bx^{2}-\frac{1}{12}(X+Y)^{2}+o(X^{2}+Y^{2}).
\end{equation*}
In this case $b$ is positive, thus this polynomial is negative and the origin is a maximum point.

\bibliography{FUbiblio}
\bibliographystyle{plain} 

\end{document}